\documentstyle[aps,preprint,tighten,epsf]{revtex}

\renewcommand{\i}{\text{i}}                
\newcommand{\e}{\mbox{e}}                  
\newcommand{\tr}{\mbox{Tr}}                
\renewcommand{\d}{\mbox{d}}                
\newcommand{\D}{\mbox{D}}                  

\begin{document}

\draft
\title{Finite density QCD
 with chiral invariant four-fermion interactions} 
\author{Michael Chavel\thanks{E-mail: chavel@uiuc.edu}}
\address{Department of Physics, University of Illinois at Urbana,\\
1110 W. Green St, Urbana, IL 61801-3080}
\date{April 20, 1997}
\maketitle

\begin{abstract}
A mean field analysis of finite density QCD is presented including the
effects of additional chiral invariant four-fermion interactions.  A
lattice regularization is used with $N_f\!=\!4$ flavors of staggered
fermions.  The use of the four-fermion coupling as an improved
extrapolation parameter over the bare quark mass in Monte Carlo
simulations is discussed.  Particular attention is given to the
structure of the phase diagram and the order of the chiral phase
transition.  At zero gauge coupling, the model reduces to a
Nambu--Jona-Lasinio model.  In this limit the chiral phase transition
is found to be second-order near the zero-density critical point and
otherwise first-order.  In the strong gauge coupling limit a
first-order chiral phase transition is found.  In this limit the
additional four-fermion interactions do not qualitatively change the
physics.  The results agree with previous studies of QCD as the
four-fermion coupling vanishes.
\end{abstract}

\section{Introduction}
\label{sec:intro}

Monte Carlo simulations of lattice gauge theories with dynamical
fermions become very computationally intensive at small fermion masses
due to the zero mass singularity of the Dirac operator.  In order to
study theories with light fermions it becomes necessary to carry
out simulations with relatively large bare masses, and then extrapolate
back to lower values.  In studying models with chiral
symmetry or dynamical chiral symmetry breaking it would be
desirable to simulate the models directly in the chiral limit, without
having to introduce any bare mass terms which explicitly break the 
symmetry.

A new approach to this problem has been
suggested\cite{Brower95,KogutS96,BarbourMK96} which does not
explicitly break chiral symmetry.  The idea is to add additional terms
to the lattice action, corresponding to chiral invariant four-fermion
interactions.  In the continuum limit, as the lattice spacing reduces
to zero, these additional terms are irrelevant\cite{WilsonK74} (in the
renormalization group sense).  However, at finite lattice spacings
they favor the generation of a dynamical fermion mass.  This dynamical
mass removes the singularity in the Dirac operator allowing for
simulations with exactly zero bare fermion
masses\cite{KogutS96,BarbourMK96}.

One particular area where this may soon lead to significant progress
is in the study of QCD at finite density.  It is generally expected
that at some density greater than that of normal nuclear matter there
is a transition to a quark-gluon plasma and chiral symmetry is
restored.  The nature and order of this transition may be an important
factor in cosmological models, neutron stars, and future heavy ion
collision experiments.
Mean field calculations at both
infinite\cite{DamgaardHK85,BarbourBDKMSW86} and
strong\cite{BilicKR92,BilicDP92} gauge coupling indicate that the transition
is first-order at low temperatures.
Unfortunately, Monte Carlo simulations of QCD at non-zero density have
not been able to give a physically accurate description of the
transition\cite{BarbourBDKMSW86,BarbourKM96}.

Quenched simulations of QCD using the grand canonical
formulation\cite{BarbourBDKMSW86} show the transition at a value of
the chemical potential equal to one half the pion mass, vanishing as
the bare quark mass $m_0\to 0$.  These results are now
understood\cite{quenched_QCD} to correspond to the $N_f\!\to\!0$ limit
of a theory of $N_f$ quarks with regular action and $N_f$ quarks with
the conjugate action, and are therefore unrelated to QCD.

Unquenched Monte Carlo calculations have to deal with a complex
fermionic determinant due to the introduction of the chemical
potential.  Hence, conventional Monte Carlo algorithms can not be
applied.  Some results have been obtained at infinite gauge coupling
by representing the partition function in terms of monomers, dimers,
and baryonic loops\cite{KarschM89}, and also on small lattices, by 
using a spectral density method\cite{Gocksch88}.  The
results are consistent with mean field calculations.  However, it is
not clear how to extend these methods to intermediate or weak gauge
coupling.

The evaluation of the complex determinant can also be avoided by the
fugacity expansion method of Refs.~\cite{Gibbs86,Barbour90}.  However,
the proper physics of chiral symmetry breaking, with a massless pion,
has not yet been achieved using finite quark masses with this
method\cite{BarbourKM96}.  Simulations of zero mass QCD with
additional chiral four-fermion interactions, called
$\chi$QCD\cite{BarbourMK96}, may be able to overcome this problem.  In
support of this approach, this paper presents a mean field analysis
of $\chi$QCD at non-zero density.  The calculations will be
restricted to the two limiting cases of zero and infinite gauge
coupling.  No restrictions are placed the strength of the four fermion
interactions.  Although in applications approximating QCD they should
be chosen relatively weak.

\section{The $\chi$QCD Model}
\label{sec:chiqcd}

The $\chi$QCD action, using the staggered fermion\cite{staggered}
formulation, is
\begin{eqnarray}
\label{Slat}
S & = & S_f + S_g, \nonumber \\
S_f & = & {N_f\over8}\gamma\sum_{\tilde x}
(\sigma^2(\tilde x)+\pi^2(\tilde x)) \nonumber \\
& \ & + \sum_{a=1}^{N_f/4}\biggl[\sum_{x,y}\bar\chi_i^a(x){\cal
M}^{ij}(x,y)\chi_j^a(y)
+\textstyle{1\over16}\displaystyle\sum_x\bar\chi_i^a(x)\chi_i^a(x)
\!\sum_{<\tilde x,x>}\bigl(\sigma(\tilde x)+\i\varepsilon(x)
\pi(\tilde x)\Bigr)\biggr], \nonumber \\
S_g & = & -\beta\sum_{\Box}[1-\frac{1}{N}{\rm Re}({\rm Tr}_{\Box} UUUU)].
\end{eqnarray}
$\bar\chi_i^a$ and $\chi_i^a$ are complex Grassmannian fields defined
on each site of the lattice.  Each $\chi_i^a$ represents four
degenerate flavors of fermions, so the sum over the flavor index $a$
runs from $1$ to $N_f/4$.  We are interested in studying QCD\@,
however, the calculations can be done for any SU($N$) gauge group.
Considering this more general case, the color indices $i$,$j$ run from
$1$ to $N$.

The fermion hopping matrix ${\cal M}^{ij}(x,y)$ is given by
\begin{equation}
\label{hop}
{\cal M}^{ij}(x,y)={1\over2}\sum_{\nu=0}^3 
\Bigl[f_{\nu}(x)U_{\nu}^{ij}(x)\delta_{y,x+\hat\nu}
-f_{\nu}^{-1}(x)U_{\nu}^{ij}(x-\hat\nu)
\delta_{y,x-\hat\nu}\Bigr]+m_0\delta_{y,x},
\end{equation}
with
\begin{equation}
f_{\nu}(x) \equiv 
	\left\{
              \begin{array}{ll}
                   \exp(\mu)   & \ \nu=0\\
                   \eta_{\nu}(x) & \ \nu=1,2,3
              \end{array}
	\right. . 
\end{equation}
Here $\mu$ is the chemical
potential\cite{HasenfratzK83,BarbourBDKMSW86} measured in units of the
lattice spacing.  The $\eta_\nu(x)$ are the conventional
Kawamoto-Smit\cite{KawamotoS81} phases $(-1)^{x_0+\cdots+x_{\nu-1}}$.
$m_0$ is the bare (current) mass of the quarks which can be set to
zero, but is included here for completeness.  

The auxiliary fields $\sigma$ and $\pi$ have been introduced to
separate the four-fermion interactions\cite{JolicoeurMP86,HandsKK93}.
They are defined on the dual lattice.  The symbol $<\! \tilde x,x \!>$
represents the $16$ dual sites $\tilde x$ adjacent to the direct
lattice site $x$. $\varepsilon(x)$ represents the alternating phase
$(-1)^{x_0+x_1+x_2+x_3}$.  After integrating over the auxiliary
fields, the lattice model approximates the continuum theory
\begin{equation}
\label{Lcont}
{\cal L}=\bar\psi(D{\!\!\!\! /}\,+\mu\gamma_0+m_0)\psi-
\textstyle{G\over2{N_f/4}}[(\bar\psi\psi)^2
-(\bar\psi\gamma_5\psi)^2] - {1\over 2g^2}\tr(F_{\mu\nu}F^{\mu\nu}),
\end{equation}
with $G\!=\!1/\gamma$ and $2N/g^2\!=\!\beta$.  (The spinor,
color and flavor indices have been suppressed.)

In the chiral limit ($m_0\!=\!0$), the continuum theory (\ref{Lcont})
is invariant under the global U(1) chiral symmetry
\begin{equation}
\bar\psi\mapsto\bar\psi\e^{\i\gamma_5\theta}, \qquad
\psi\mapsto\e^{\i\gamma_5\theta}\psi.
\end{equation}
In the language of the lattice model this translates to
\begin{equation}
\label{transform}
 \bar\chi^a \mapsto \bar\chi^a
\e^{\i\theta\varepsilon(x)} \;,\;\;\; \chi^a \mapsto
\e^{\i\theta\varepsilon(x)}\chi^a \;,\;\;\; \mbox{and}\;\;\;
\left(
\begin{array}{l}
	\sigma \\
	\pi \\
\end{array}
\right) \mapsto 
\left(
\begin{array}{ll}
	\cos(2\theta) & \sin(2\theta) \\
	-\sin(2\theta) & \cos(2\theta) \\
\end{array}
\right)
\!\!\left(
\begin{array}{l}
	\sigma \\
	\pi \\
\end{array}
\right).
\end{equation}
The U(1) chiral symmetry is spontaneously broken in the low density
hadron phase through the dynamics of the color gauge fields.  The
chiral condensate
\begin{equation}
\langle\bar\chi^a\chi^a\rangle = {N_f\over 4}\gamma\langle\sigma\rangle,
\end{equation}
is nonzero in the broken phase, and serves as the order parameter for
the chiral phase transition.

\section{Zero Gauge Coupling --- NJL Limit}
\label{sec:NJL}

As the gauge coupling vanishes $\chi$QCD reduces to a lattice
Nambu--Jona-Lasinio (NJL) model\cite{NJL}. In Ref.~\cite{HandsKK95}
the three dimensional version of this model was studied at non-zero
chemical potential in a $1/N_f$ expansion.  Presented here are results
for four dimensions.  Of particular interest is the structure of the
finite density phase diagram and the order of the chiral phase
transition.

When $\beta\!\to\!\infty$ the SU($N$) gauge variables are frozen to
$U_{\nu}^{ij}(x)\!=\!\delta^{ij}$ at each link of the lattice.
Gaussian integration over the fermion fields leaves the following, bosonic,
partition function
\begin{equation}
\label{NJLZ}
Z = \int\D\sigma \D\pi
\ \exp\biggl\{-\displaystyle{N_f\over8}\gamma\sum_{\tilde x}
(\sigma^2(\tilde x)+\pi^2(\tilde x)) + {N\!\cdot\!N_f\over
4}\tr\ln[S_F^{-1}]\biggr\}, 
\end{equation}
where
\begin{equation}
S_F^{-1}(x,y) = {1\over2}\sum_{\nu=0}^3\
\Bigl[f_\nu(x)\delta_{y,x+\hat\nu}-f_\nu^{-1}(x)\delta_{y,x-\hat\nu}\Bigr]
+m_0+\!\textstyle{1\over 16}\!\displaystyle
\sum_{<\tilde x,x>}(\sigma(\tilde x)+\i\varepsilon(x)\pi(\tilde x)).
\end{equation}

The mean field, saddle point solution $\{\bar\sigma, \bar\pi\}$
(equivalent to the first order of a large $N_f$ expansion) is given by
the ``gap equation''
\begin{equation}
\label{NJLgap}
\left(
	\begin{array}{c}
		\bar\sigma \\
		\bar\pi \\
	\end{array}
\right)\gamma =
\left(
	\begin{array}{c}
		\bar\sigma + m_0 \\
		\bar\pi \\
	\end{array}
\right)N\tr[S_F],
\end{equation}
where (in the infinite volume limit)
\begin{eqnarray}
\label{NJLtr}
\tr[S_F] &=& 16\int_{-{\pi\over 2}}^{\pi\over 2}{\d^4 k \over (2\pi)^4}
{1\over\textstyle{1\over 2}\bigl(1-\cos2k_0\cosh2\mu - \i
\sin2k_0\sinh2\mu\bigr)+\displaystyle\sum_{\nu=1,2,3}\sin^2
k_{\nu}+m^2}, \nonumber \\
m^2 &=& (\bar\sigma + m_0)^2 + \bar\pi^2.
\end{eqnarray}  
Here $m$ is the effective (dynamical) mass of the quarks.  If the bare
quark mass $m_0$ is non-zero, chiral symmetry is broken explicitly. The
unique solution is then identified with $\bar\sigma\!>\!0$ and
$\bar\pi\!=\!0$.  In the chiral limit the solution is invariant under
the transformation (\ref{transform}), and so we are free to choose
$\bar\pi=0$.  This is the case of interest, which we will now discuss
for $N\!=\!3$.

The gap equation is easily evaluated numerically on a finite size
lattice by replacing the momentum integrations with discrete
sums\cite{HandsKK95}.  A $(48)^4$ Euclidean lattice was found to be
sufficiently large to eliminate finite-size effects.
Figure \ref{fig:NJLMCdata} shows the resulting behavior of
$\bar\sigma$ at zero chemical potential.  The other data points show
(zero mass) measurements of $\sqrt{\sigma^2+\pi^2}$ from Monte Carlo
simulations of $\chi$QCD in the $\beta\!\to\!\infty$
limit\cite{KogutLMB97}.  The mean field phase transition is at
$\gamma_{\mbox{c}}\!=\!1.86$.  This is also in agreement with
Monte Carlo simulations of the pure NJL model\cite{KimKK94}, where
the transition was found at $\gamma_{\mbox{c}}\!=\!1.81$
($\gamma_{\mbox{c}}/N\!\Leftrightarrow\!\beta_{\mbox{c}}$ in their
notation).

\begin{figure}
\begin{center}
\leavevmode
\epsfxsize=4.3in
\epsfbox{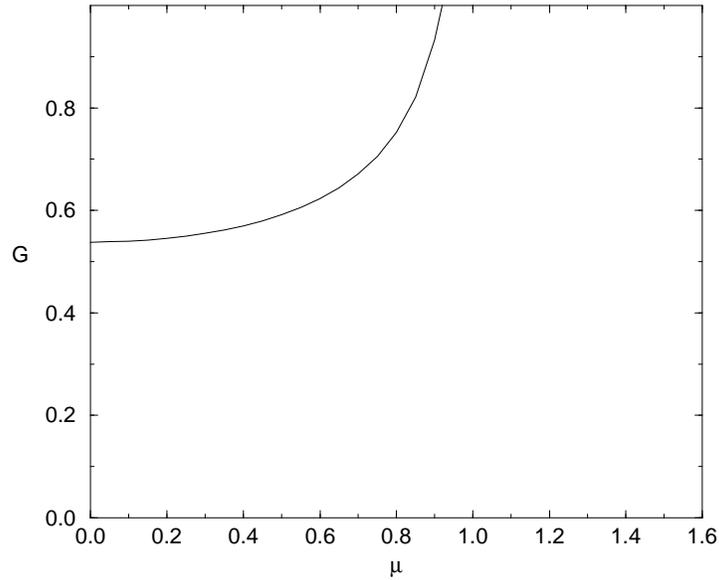}
\caption{The mean field phase diagram of $\chi$QCD at
$\beta\!\to\!\infty$.  Plotted here is the critical line separating
the dynamically broken phase, at large $G$, from the chiral symmetric
phase, at small $G$.  $G\!=\!1/\gamma$.}
\label{fig:NJLphase}
\end{center}
\end{figure}

At non-zero chemical potential the chiral phase transition is given by
setting $\bar\sigma\!=\!0$.  The gap equation then produces the
critical line shown in the phase diagram of Fig.~\ref{fig:NJLphase}.
The upper left section of the phase diagram corresponds to the
spontaneously broken symmetry phase, with $\bar\sigma\!>\!0$.  The
remainder of the diagram represents the chiral symmetric phase, with
$\bar\sigma\!=\!0$.

\begin{figure}
\begin{center}
\leavevmode
\epsfxsize=4.3in
\epsfbox{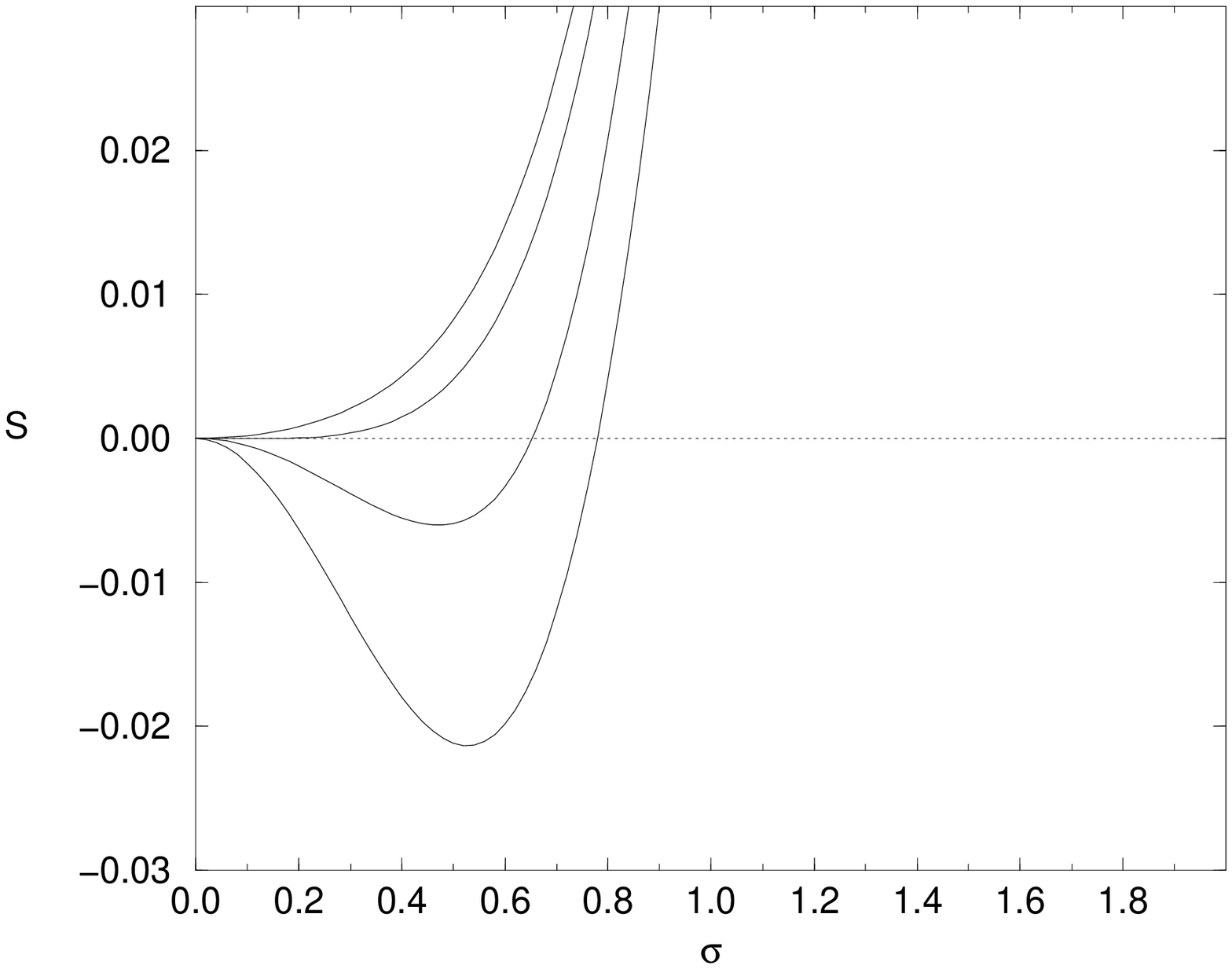}
\caption{The $\beta\!\to\!\infty$ effective action of $\chi$QCD plotted as a
function of $\bar\sigma$, for $\gamma\!=\!1.5$. The different lines are
for $\mu\!=\!0$, $\mu\!=\!.60$, $\mu\!=\!.69$, and $\mu\!=\!.72$, from
bottom to top.}
\label{fig:NJLSeff1.5}
\end{center}
\end{figure}

\begin{figure}
\begin{center}
\leavevmode
\epsfxsize=4.3in
\epsfbox{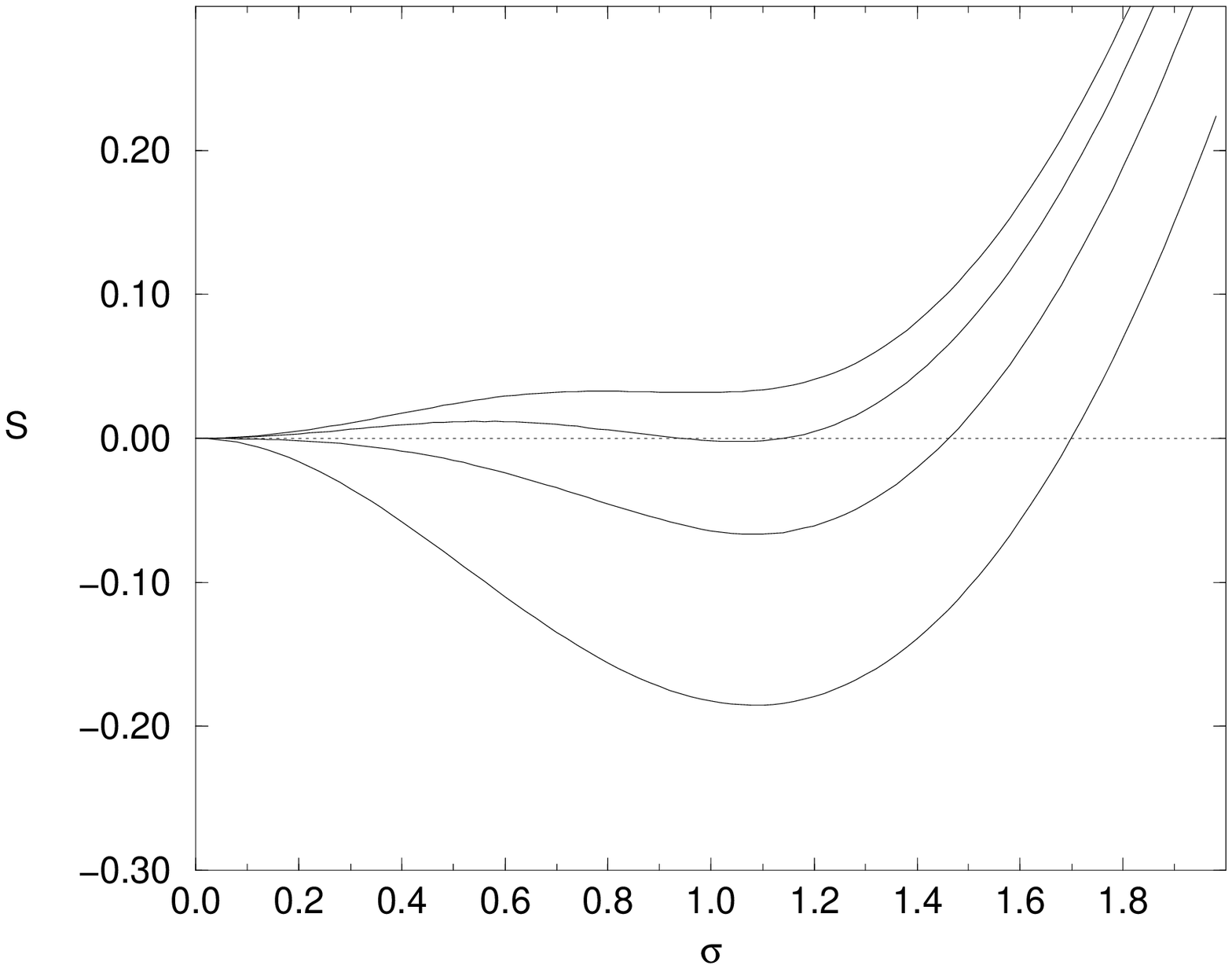}
\caption{The $\beta\!\to\!\infty$ effective action of $\chi$QCD
plotted as a function of $\bar\sigma$, for $\gamma\!=\!1$. The three
different lines are for $\mu\!=\!0$, $\mu\!=\!.90$, $\mu\!=\!.97$, and
$\mu\!=\!1.0$ from bottom to top.}
\label{fig:NJLSeff1}
\end{center}
\end{figure}

As an alternative to solving (\ref{NJLgap}) the minimum energy
solution can be found by a direct evaluation of the action.  This
also allows for a clearer determination of the order of the chiral phase
transition.  Substituting the uniform ansatz $\{\bar\sigma,0\}$ into
the action and Fourier transforming one finds
\begin{eqnarray}
\bar S=\textstyle{1\over 2}\gamma\bar\sigma^2 -
3\!\cdot\!16\displaystyle \int_{-{\pi\over2}}^{\pi\over 2}{\d^4 k \over
(2\pi)^4} \ln\biggl[\textstyle{1\over 2}\bigl(1 - \cos2k_0\cosh2\mu &-& \i
\sin2k_0\sinh2\mu\bigr) \nonumber \\
&+& \displaystyle\sum_{\nu=1,2,3}\sin^2 k_{\nu}+\bar\sigma^2\biggr].
\end{eqnarray}
Figure~\ref{fig:NJLSeff1} contains plots of $\bar S$ as a function of
$\bar\sigma$ at $\gamma\!=\!1$.  Each line corresponds to a different
value of the chemical potential.  The plots have been rescaled so that
$\bar S$ is zero at $\bar\sigma\!=\!0$.  The figure shows a clear
first-order (discontinuous) transition as the chemical potential is
increased from zero.  However, at larger values of $\gamma$ the
transition becomes continuous, as shown in Fig.~\ref{fig:NJLSeff1.5}
for $\gamma\!=\!1.5$.  The chiral transition remains continuous as the
critical line is followed towards $\mu\!=\!0$.  This appears
consistent with the second-order transition found in the $1/N_f$
expansion at zero-density
\cite{KimKK94}.\footnote{
Monte Carlo calculations\cite{KimKK94} indicate that the
theory is in fact trivial, exhibiting logarithmic scaling corrections
to mean field exponents.}
What is perhaps unanticipated is that the second-order transition
actually extends into the interior of the phase diagram for small
values of the chemical potential.

\begin{figure}
\begin{center}
\leavevmode
\epsfxsize=4.3in
\epsfbox{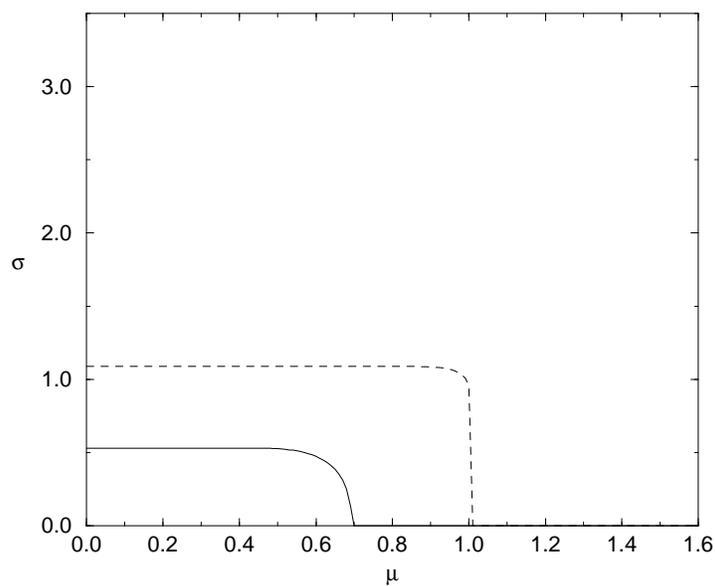}
\caption{$\bar\sigma$ as a function
 of the chemical potential at $\beta\to\infty$ for
$\gamma\!=\!1.5$ (solid line) and $\gamma\!=\!1$ (dashed line).}
\label{fig:NJLsigma}
\end{center}
\end{figure}

\begin{figure}
\begin{center}
\leavevmode
\epsfxsize=4.3in
\epsfbox{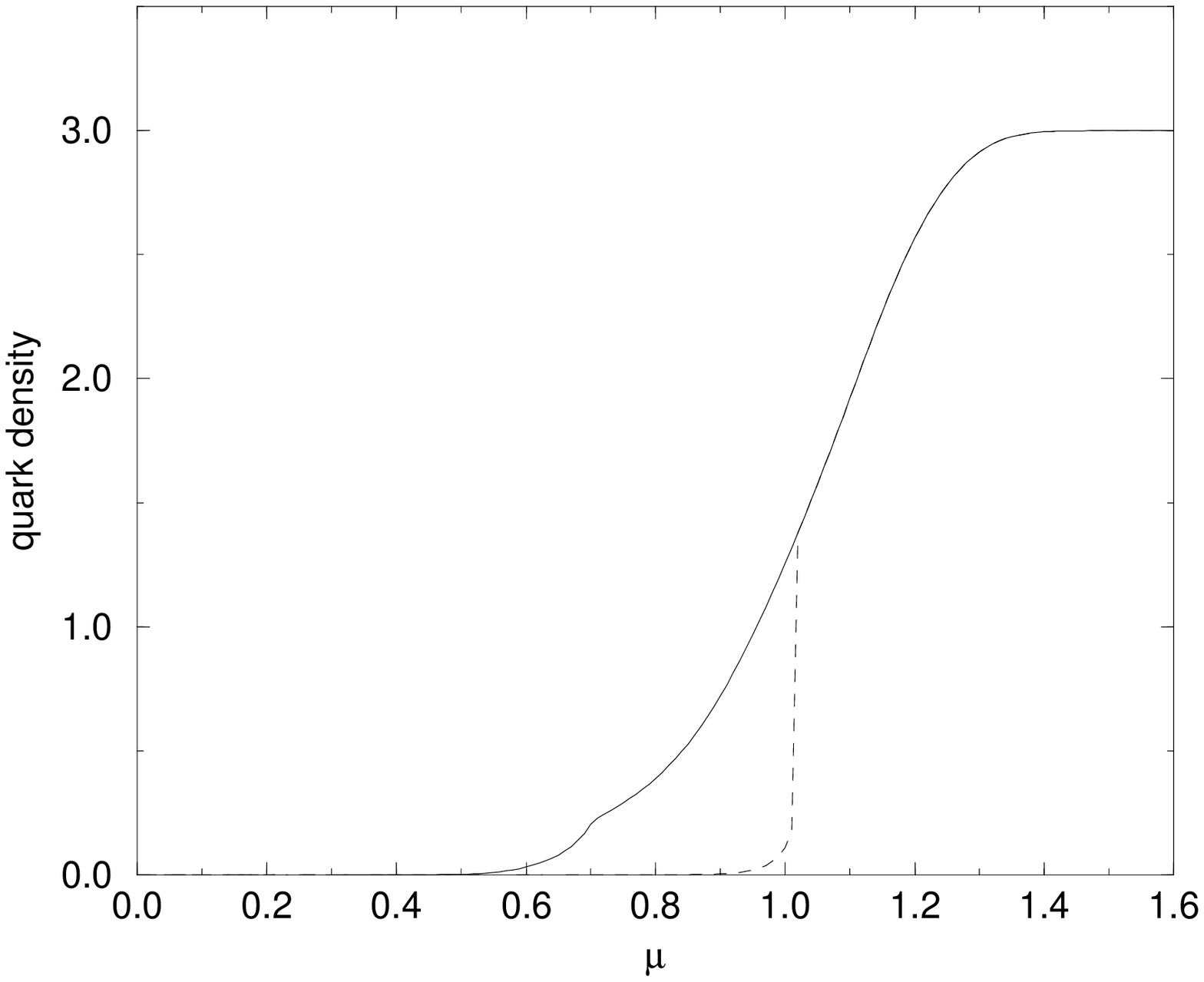}
\caption{Plots of the quark density at $\beta\to\infty$ for
$\gamma\!=\!1.5$ (solid line) and $\gamma\!=\!1$ (dashed line), as a
function of the chemical potential.}
\label{fig:NJLnum}
\end{center}
\end{figure}

\begin{figure}
\begin{center}
\leavevmode
\epsfxsize=4.3in
\epsfbox{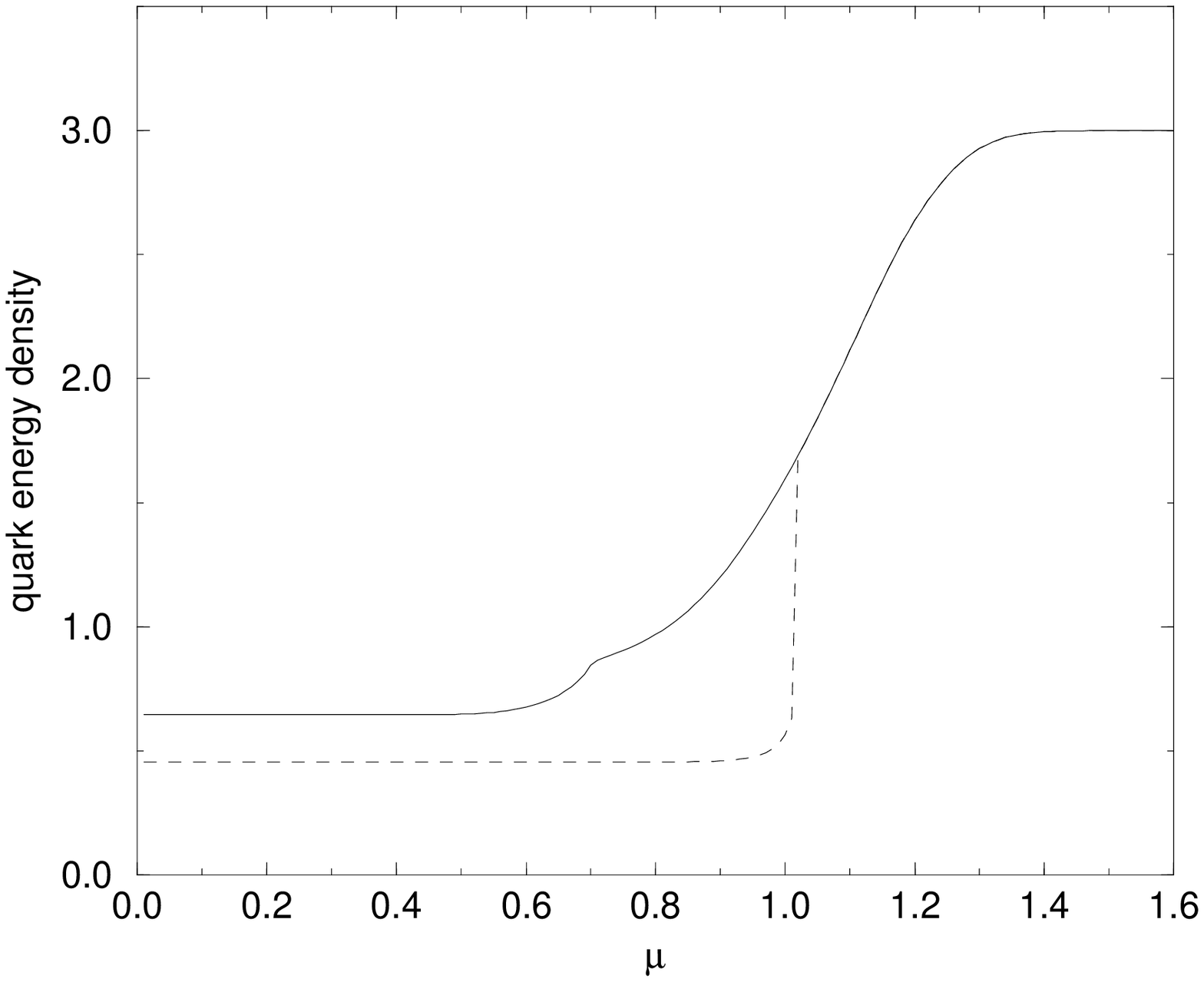}
\caption{Plots of the quark energy density at
$\beta\to\infty$ for $\gamma\!=\!1.5$ (solid line) and $\gamma\!=\!1$
(dashed line), as a function of the chemical potential.}
\label{fig:NJLenergy}
\end{center}
\end{figure}

The behavior of $\bar\sigma$ as a function of the chemical potential
is plotted in Fig.~\ref{fig:NJLsigma}, for the fixed values of
$\gamma\!=\!1$ and $1.5$.  Other quantities of interest include the quark
number density and quark energy density.
\begin{eqnarray}
\langle\psi^{\dagger}\!\psi\rangle 
&=& -{\partial\ln Z\over\partial\mu} \nonumber \\
&=& -N\tr\biggl[{\partial S_F^{-1}\over\partial\mu}S_F\biggr]
\nonumber \\
&=& 3\!\cdot\! 16\int_{-{\pi\over 2}}^{\pi\over 2}{\d^4 k \over (2\pi)^4}
{\textstyle{1\over
2}\bigl(\cos2k_0\sinh2\mu + \i\sin2k_0\cosh2\mu\bigr)
\over\textstyle{1\over 2}\bigl(1-\cos2k_0\cosh2\mu - \i
\sin2k_0\sinh2\mu\bigr)+\displaystyle\sum_{\nu=1,2,3}\sin^2 k_{\nu}
 + \bar\sigma^2}, \nonumber \\
\langle\psi^{\dagger}\!\partial_0\psi\rangle
&=& N\tr\bigl[(S_F^{-1})^0 S_F\bigr] \nonumber \\
&=& 3\!\cdot\!16\int_{-{\pi\over 2}}^{\pi\over 2}{\d^4 k \over (2\pi)^4}
{\textstyle{1\over 2}\bigl(1-\cos2k_0\cosh2\mu - \i
\sin2k_0\sinh2\mu\bigr)\over
\textstyle{1\over 2}\bigl(1-\cos2k_0\cosh2\mu - \i
\sin2k_0\sinh2\mu\bigr)+\displaystyle\sum_{\nu=1,2,3}\sin^2 k_{\nu}
+ \bar\sigma^2}.
\end{eqnarray}
These are plotted in Fig.~\ref{fig:NJLnum} and
Fig.~\ref{fig:NJLenergy}, respectively, for the values of
$\gamma\!=\!1$ and $1.5$.  As expected, the quark density is
essentially zero in the broken phase.  As the chemical potential is
increased towards the critical value separating the two phases,
roughly equal to $m\!=\!\bar\sigma$, the order parameter begins to
decrease and the density rises.  At the chiral transition the quark
mass abruptly vanishes.  Hence, the quark density abruptly increases.
In the symmetric phase, the number density increases $\sim\mu^3$ at
low densities, which is consistent with a non-interacting quark gas.
As the lattice becomes filled the density eventually saturates at the
maximum value allowed by the Pauli exclusion principle, of $N\!=\!3$.

\section{The Strong Gauge Coupling Limit}
\label{sec:strong}

In the previous section we studied the case where the gauge
coupling vanished.  However, in using $\chi$QCD to approximate QCD one
is really interested in the case where the four-fermion interactions are
very weak compared to the gauge interactions; naively when $G\!\ll
g$.  In this section we will study the strong gauge coupling limit of
$g\!\to\!\infty$.  Similar infinite gauge coupling 
calculations have been carried out in
\cite{Kluberg-SternMP83}, as well as in
\cite{DamgaardHK85,BarbourBDKMSW86} at finite density.
Some of the same techniques will be used here with suitable
modifications.  

Begin by defining the following ``meson'' and ``baryon'' fields:
\begin{eqnarray}
\label{MBdef}
M^{ab}(x) &\equiv& \bar\chi_i^a(x) \chi_i^b(x), \nonumber \\
\bar B^{ab\ldots c}(x) &\equiv& {1\over N!}\epsilon_{ij\ldots k}
\bar\chi_i^a(x)\bar\chi_j^b(x)\ldots\bar\chi_k^c(x), \nonumber \\
B^{ab\ldots c}(x) &\equiv& {1\over N!}\epsilon_{ij\ldots k}
\chi_i^a(x)\chi_j^b(x)\ldots\chi_k^c(x).
\end{eqnarray}
The meson fields commute for any value of N, while the baryon fields
will anticommute (commute) for $N$ odd (even). 

At infinite gauge coupling ($\beta\!=\!0$) the gauge term in the action
$S_g$ is zero. Therefore, the gauge variables on each link are
completely independent.  Integrating over the SU($N$) gauge group at
each link yields\footnote{
Equation (\ref{Zexpan}) is valid only for
$N\!\geq\!3$.  Similar expressions for the U(1) and SU(2) gauge
groups can be found in Ref.~\cite{Kluberg-SternMP83}}
\begin{eqnarray}
\label{Zexpan}
Z &=& \int\! \D\bar\chi \D\chi \D\sigma \D\pi \nonumber \\
&\times& \exp\biggl\{-\displaystyle{N_f\over8}\gamma\sum_{\tilde x}
(\sigma^2(\tilde x)+\pi^2(\tilde x)) - \displaystyle\sum_{a=1}^{N_f/4}
\sum_x\bar\chi_i^a(x)\chi_i^a(x)\biggl[ m_0 +
\textstyle{1\over 16}\!\displaystyle\sum_{<\tilde x,x>}(\sigma(\tilde
x)+\i\varepsilon(x)\pi(\tilde x))\biggr]\biggr\} \nonumber \\
&\times& \prod_x \!\prod_{\nu=0}^3 \biggl[ \ 1 + {1 \over 4N}\tr_f[M(x)
M(x\!+\!\tilde\nu)] \nonumber \\
& \ & \quad  + {1\over 32N(N^2\!-\!1)}
\biggl(\tr_f[M(x)M(x\!+\!\tilde\nu)]^2 +
N\tr_f[M(x)M(x\!+\!\tilde\nu)M(x)M(x\!+\!\tilde\nu)]\biggr) +
\ldots  \nonumber \\
& \ & \quad  - \ {(-1)^{N(N+1)/2}\over2^N}
\biggl(f_{\nu}^N(x)
\tr_f[\bar B(x)B(x\!+\!\tilde\nu)] + (-1)^Nf_{\nu}^{-N}(x)
\tr_f[\bar B(x\!+\!\tilde\nu)B(x)]\biggr) \ \biggr]. 
\end{eqnarray}
Here $\tr_f$ is the trace over flavor indices only.  In writing
(\ref{Zexpan}), the full expression for the meson contributions has
been truncated.  (In general the series will terminate after terms of
order $O(M^{NN_f/2})$ due to the Grassmannian nature of the $\chi_i^a$
fields.)  For the remainder of the calculation we only need to retain
the first meson-meson term.  As shown in \cite{Kluberg-SternMP83} this
is consistent with a systematic expansion in $1/d$, where $d$ is the
number of space-time dimensions. 

Exponentiating the product at each lattice site yields the
following action:
\begin{eqnarray}
S = \textstyle{N_f\over 8}\gamma &\displaystyle\sum_{\tilde x}&
(\sigma^2(\tilde x)+\pi^2(\tilde x)) + 
\sum_x\tr_f[M(x)]\biggl( m_0 +
\textstyle{1\over 16}\displaystyle
\sum_{<\tilde x,x>}(\sigma(\tilde x)+\i\varepsilon(x)
\pi(\tilde x))\biggr) \nonumber \\
&-& \sum_{x,y} \biggl(\textstyle{1\over
2}\tr_f[M(x)V(x,y)M(y)]+\tr_f[\bar B(x)V_B(x,y)B(y)]\biggr),
\end{eqnarray}
with
\begin{eqnarray}
\label{Vdefs}
V(x,y) & \equiv & {1 \over 4N}\sum_{\nu=0}^3\bigl(\delta_{y,x+\hat\nu} + 
\delta_{y,x-\hat\nu}\bigr), \\
V_B(x,y) & \equiv & {(-1)^{N(N+1)/2} \over 2^N}
\sum_{\nu=0}^3\bigl(f_{\nu}^N(x)\delta_{y,x+\hat\nu}+
(-1)^Nf_{\nu}^{-N}\delta_{y,x-\hat\nu}\Bigr).
\end{eqnarray}
Note that all the dependence on the chemical potential is now
contained in $V_B(x,y)$.

We want to construct an effective action by
integrating out the staggered fermion fields, as was done in the
pure NJL case.  To facilitate this we first linearize the action in the
meson and baryon terms using the following identities:
\begin{eqnarray}
& \exp & \biggl\{\displaystyle\sum_{x,y}\textstyle{1\over
2}\tr_f[M(x)V(x,y)M(y)]\biggr\} \nonumber \\
& \ & \quad  = \det[V]\displaystyle\int\! \D\rho \;
\exp\biggl\{-\displaystyle\sum_{x,y}\textstyle{1\over 2}
\tr_f[\rho(x)V^{-1}(x,y)\rho(y)]-
\displaystyle\sum_x\rho^{ab}(x)M^{ab}(x)\biggr\}, \\
& \exp & \biggl\{\displaystyle\sum_{x,y}\tr_f[\bar
B(x)V_B(x,y)B(y)]\biggr\} \nonumber \\
& \ & \quad = \det[V_B]\int\! \D\bar b \D b \;
\exp\biggl\{-\displaystyle\sum_{x,y}
\tr_f[\bar b(x)V_B^{-1}(x,y)b(y)]\biggr\} \nonumber \\
& \ & \qquad\qquad\qquad\qquad \times
\exp\biggl\{-\displaystyle\sum_x\bigl(\bar b^{ab\ldots
c}(x)B^{ab\ldots c}(x) + \bar B^{ab\ldots c}(x)b^{ab\ldots
c}(x)\bigr)\biggr\}.
\label{B_identity}
\end{eqnarray}
In writing the second identity we have taken $N$ to be odd, as it is
for QCD.  In either case, the final expression for the effective
action (\ref{S_eff}) will be correct for all $N\geq 3$.\footnote{
For $N$ even $\det[V_B]$ is replaced by $\det[V_B^{-1}]$ in
(\ref{B_identity}).  Integration over $\bar b, b$ in
(\ref{Zbb}) then gives $\int\!D\!\bar b D\! b \,\bar b b\,
\e^{-\bar b V_B^{-1}b} = \det[V_B^2]$.  Combining all the
determinants eventually yields equation (\ref{S_eff}).}

Let us now restrict the remainder of the calculation to $N_f\!=\!4$
flavors which is technically much easier to handle.  (Unlike the high
temperature transition (at low density), the finite density transition
should not have a large flavor dependence at low
temperatures\cite{BilicKR92}.)  Integrating over $\bar\chi_i^a$ and
$\chi_i^a$ we then find
\begin{eqnarray}
\label{Zbb}
Z &=& \int\! \D\sigma \D\pi \D\rho \D\bar b \D b
\ \exp\biggl\{-{1\over 8}\gamma\displaystyle\sum_{\tilde x}
(\sigma^2(\tilde x)+\pi^2(\tilde x))\biggr\}\det[V]\det[V_B] \nonumber \\
& \ & \quad \times \det\biggl[\biggl(\rho(x)+m_0+\textstyle{1\over 16}\!
\displaystyle\sum_{<\tilde x,x>}(\sigma(\tilde x)+\i\varepsilon(x)\pi(\tilde
x))\biggr)^N \!+ (-1)^{N(N-1) \over 2}\bar b(x)b(x)\biggr] \nonumber \\
& \ & \quad \times \exp\biggl\{ -\displaystyle\sum_{x,y}
\biggl[\textstyle{1\over2}\rho(x)V^{-1}(x,y)\rho(y)+
\bar b(x)V_B^{-1}(x,y)b(y)\biggr]\biggr\}.
\end{eqnarray}
Since $\det[V]$ does not depend upon any of the fields or the chemical
potential, it can be dropped without affecting the calculation of any
physical quantities.  The remaining functional determinants can be
exponentiated up into the action,
producing an effective action
\begin{eqnarray}
\label{Seffbb1}
S_{eff} = \textstyle{1\over 2}\gamma\displaystyle\sum_{\tilde
x}(\sigma^2(\tilde x)+\pi^2(\tilde x)) &+& \sum_{x,y}\biggl(\textstyle{1\over
2}\rho(x)V^{-1}(x,y)\rho(y)+\bar b(x)V_B^{-1}(x,y)b(y)\biggr)
\nonumber \\
&-& \tr\ln\bigl[\Sigma^N(x) + (-1)^{N(N-1) \over 2}\bar b
b\bigr] -\tr\ln[V_B],
\end{eqnarray}
with
\begin{equation}
\Sigma(x) \equiv \rho(x)+m_0+\textstyle{1\over 16}\!\displaystyle
\sum_{<\tilde x,x>}\bigl(\sigma(\tilde x)+\i\varepsilon(x)\pi(\tilde x)\bigr).
\end{equation}
Using the fact that $\ln[1 + \textstyle{\bar b b\over\Sigma}] = {\bar
b b\over\Sigma}$ for the Grassmannian fields $\bar b$ and $b$,
(\ref{Seffbb1}) can be further simplified to
\begin{eqnarray}
\label{S_effbb2}
S_{eff} = \textstyle{1\over 2}\gamma\displaystyle\sum_{\tilde
x}(\sigma^2(\tilde x)+\pi^2(\tilde x)) &+&
\sum_{x,y}\biggl(\textstyle{1\over2}\rho(x)V^{-1}(x,y)
\rho(y)+\bar b(x)S_b^{-1}(x,y)b(y)\biggr) \nonumber \\
&-& \tr\ln\bigl[\Sigma^N(x)\bigr] - \tr\ln[V_B], 
\end{eqnarray}
with
\begin{equation}
\label{S_b}
S_b^{-1}(x,y) = V_B^{-1}(x,y) + {\delta(x,y)\over\Sigma(x)^N}.
\end{equation}
Integrating over $\bar b$ and $b$, which are now just Gaussian integrals, we
arrive at a final expression for the effective action at
$\beta\!=\!0$
\begin{equation}
\label{S_eff}
S_{eff} =
\textstyle{1\over 2}\gamma\displaystyle\sum_{\tilde x}
(\sigma^2(\tilde x)+\pi^2(\tilde x))
+ \sum_{x,y}\textstyle{1\over 2}\rho(x)V^{-1}(x,y)\rho(y) -
\tr\ln\bigl[\Sigma^N(x) + V_B(x,y)\bigr].
\end{equation}

Now consider a uniform (mean field) saddle point approximation to
the action.  Denote the saddle point solution which minimizes
the effective action by $\{\bar\rho, \bar\sigma, \bar\pi\}$.  Substituting
this into (\ref{S_eff}) and transforming to momentum space yields
\begin{eqnarray}
\label{S_effk}
\bar S_{eff} = \textstyle{1\over 2}\gamma(\bar\sigma^2+\bar\pi^2)
+ \textstyle{N\over4}\bar\rho^2 \nonumber - 8\displaystyle
\int_{-{\pi\over 2}}^{\pi\over 2}{\d^4 k \over (2\pi)^4}
\ln\biggl[\textstyle{1\over2}\bigl(1 &-& \cos2k_0\cosh2N\mu - \i
\sin2k_0\sinh2N\mu\bigr) \nonumber \\
&+& \displaystyle\sum_{\nu=1,2,3}\sin^2 k_{\nu} + m_B^2 \biggr],
\end{eqnarray}
with
\begin{equation}
m_B^2 = 4^{N-1}[(\bar\rho+\bar\sigma +m_0)^2 +\bar\pi^2]^N.
\end{equation}
$\bar S_{eff}$ can be evaluated by an explicit summation over discrete
momenta on a finite sized lattice, as carried out in section
\ref{sec:NJL}.
In the chiral limit ($m_0\!=\!0$) we find two distinct phases: For
$\mu\!<\!\mu_c(\gamma)$ there is a spontaneously broken phase
with\footnote{
Since the values of $\{\bar\rho, \bar\sigma, \bar\pi\}$ are
independent of $\mu$ in each phase the analytic expressions
(\ref{analytic}) can be obtained with $\mu\!=\!0$.  In which case, the
baryon terms can be safely ignored\cite{BarbourBDKMSW86} and
$V_B(x,y)$ can be set equal to zero.}
\begin{equation}
\label{analytic}
\bar\rho = \sqrt{2\over 1+{N\over 2\gamma}}\;,\;\;
\bar\sigma = {N\over 2\gamma}\bar\rho \;,\;\;
\bar\pi = 0;
\end{equation}
while for $\mu\!>\!\mu_c(\gamma)$ there is a chiral symmetric phase,
characterized by
\begin{equation}
\bar\rho = \bar\sigma = \bar\pi = 0.
\end{equation}
(The same solutions are produced by evaluating the ``gap equation'',
obtained by minimizing (\ref{S_effk}) with respect to $\{\bar\rho,
\bar\sigma, \bar\pi\}$.)
Notice that within each phase all of the order parameters are
independent of the value of the chemical potential.  Therefore, the chiral
phase transition is discontinuous (first-order) for all values of
$\gamma$.  The critical line can be determined by
setting $\bar S_{eff}(\{\bar\rho,\bar\sigma,\bar\pi\}; \mu_c) = \bar
S_{eff}(\{0,0,0\}; \mu_c)$.  This yields
\begin{equation}
\label{critline}
\mu_c(\gamma) = \textstyle{1\over2}\ln[4(2+N/\gamma)] -  
\textstyle{1\over2},
\end{equation}
in agreement with previous mean field QCD
calculations\cite{BarbourBDKMSW86,BilicDP92} as $\gamma\to\infty$.  

\begin{figure}
\begin{center}
\leavevmode
\epsfxsize=4.3in
\epsfbox{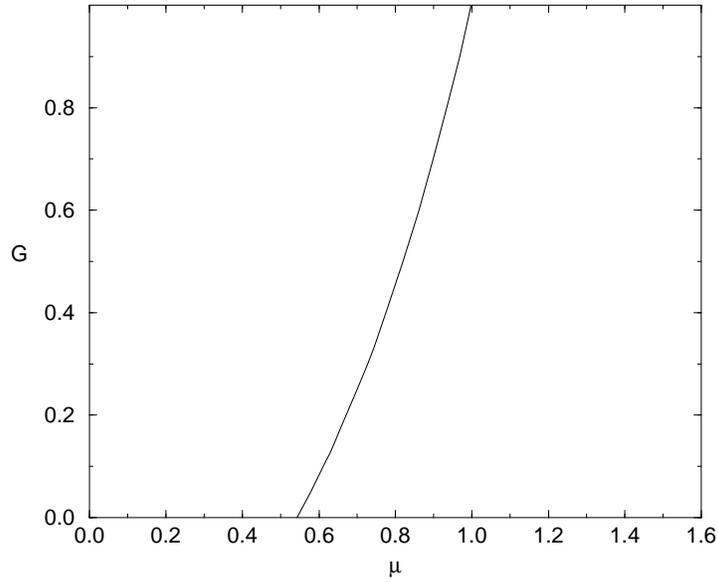}
\caption{The mean field phase diagram of $\chi$QCD at $\beta\!=\!0$.
The solid line is the critical line separating the dynamically broken
phase at small $\mu$, from the chiral symmetric phase at large $\mu$.
$G\!=\!1/\gamma$.}
\label{fig:phase}
\end{center}
\end{figure}

Let us now discuss the specific gauge group SU(3).  The phase diagram
is shown in Fig.~\ref{fig:phase}.  As expected, the values of $\mu_c$
increase monotonically as $G\!=\!1/\gamma$ increases.  Of course,
since $g\!\gg\!G$, chiral symmetry breaking is still due to the strong
gauge (color) forces as in ordinary QCD\@.  The additional
four-fermion interactions merely enhance the effect.  (This is clearly
demonstrated by comparing Fig.~\ref{fig:phase} with the phase diagram
produced in the absence of the gauge fields (Fig.~\ref{fig:NJLphase}).)
The mean field chiral condensate is given by
\begin{equation}
\langle\bar\chi\chi\rangle = \gamma\bar\sigma  = \textstyle{3\over2}\bar\rho.
\end{equation}
Values of $\bar\rho$ and $\bar\sigma$ in the broken phase are plotted
as a function of $G\!=\!1/\gamma$ in Fig.~\ref{fig:MCdata}.  Included for
comparison are values of $\sigma$ from exploratory Monte Carlo
simulations of $\chi$QCD\cite{KogutLMB97} at $\beta=0.5$ and zero chemical
potential. 

\begin{figure}
\begin{center}
\leavevmode
\epsfxsize=4.3in
\epsfbox{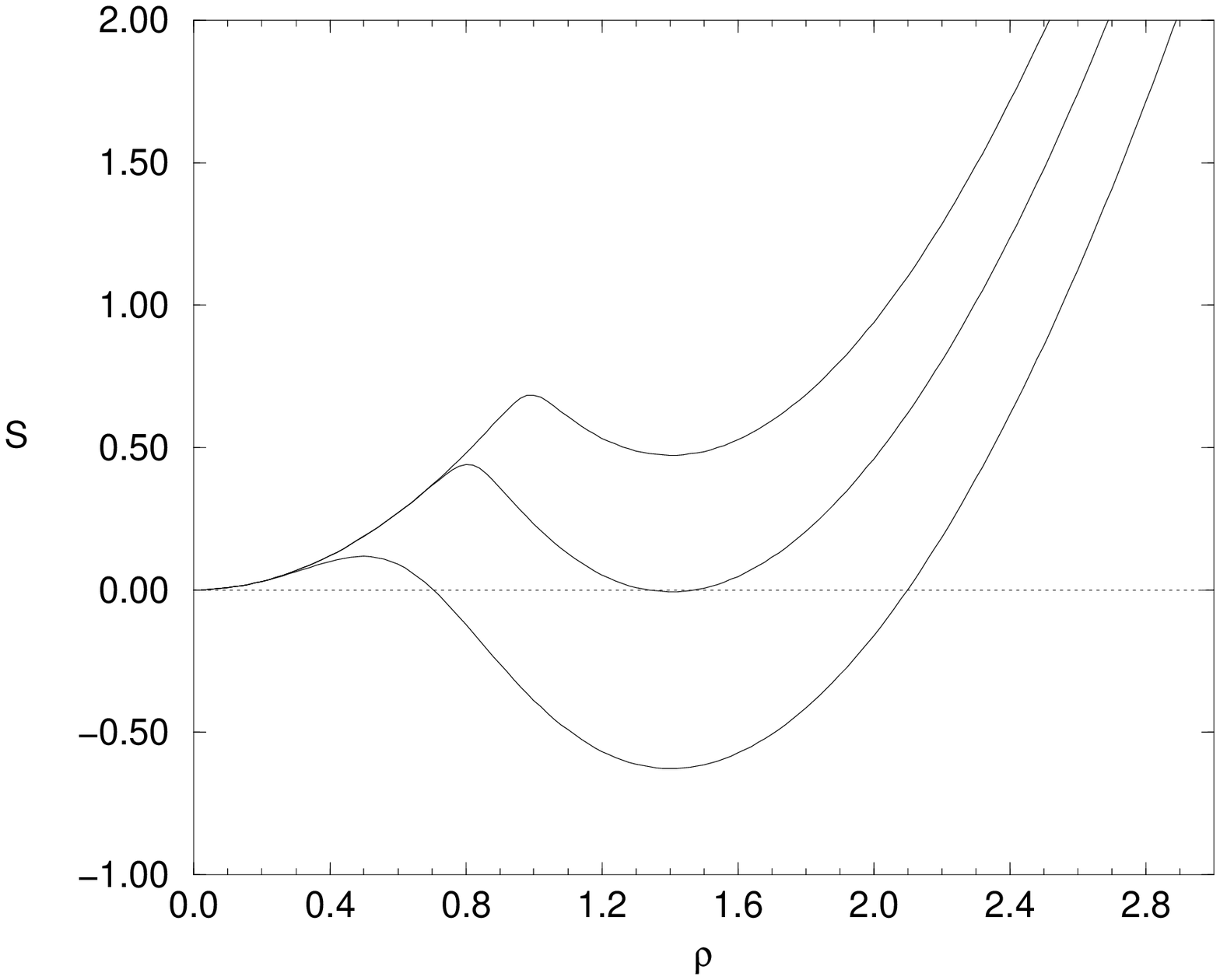}
\caption{The $\beta\!=\!0$ effective action of $\chi$QCD plotted as a
function of $\bar\rho$, for $\gamma\!\to\!\infty$. The three different
lines are for $\mu\!=\!0$, $\mu\!=\!.54$ and $\mu\!=\!.7$, from bottom
to top.}
\label{fig:Seff1}
\end{center}
\end{figure}

\begin{figure}
\begin{center}
\leavevmode
\epsfxsize=4.3in
\epsfbox{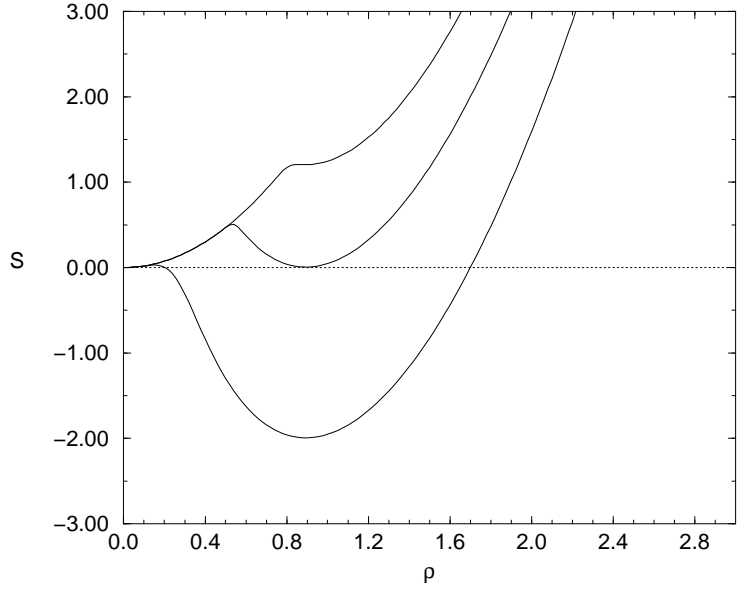}
\caption{The $\beta\!=\!0$ effective action of $\chi$QCD plotted as a
function of $\bar\rho$, at $\gamma=1$. The three different lines are for
$\mu\!=\!0$, $\mu\!=\!1.0$ and $\mu\!=\!1.4$, from bottom to top.}
\label{fig:Seff2}
\end{center}
\end{figure}

Plots of the effective action for $\gamma\!\to\!\infty$ are shown in
Fig.~\ref{fig:Seff1} at selected values of the chemical potential
$\mu$.  It should be obvious from the figure that the chiral phase
transition is indeed discontinuous, and that $\bar\rho$ does not
depend on the value of the chemical potential within the broken phase.
The value of $\mu_c\!=\!0.54$ agrees with the mean field result of
\cite{BarbourBDKMSW86} for ordinary QCD\@.  Figure
\ref{fig:Seff2} contains similar plots of the effective action
for $\gamma\!=\!1$.  Notice that the general behavior of the action
and the order of the transition remain unchanged.  (This was
explicitly checked for values down to $\gamma\!=\!.01$.)

Let us now consider the mass of the baryons in the broken phase.  By
adding baryon source terms to the original action, it is
straightforward to show that the staggered baryon propagator
$\langle\bar B(x)B(y)\rangle$ has the same pole structure as
$S_b(x,y)$, defined in (\ref{S_b}).  Defining the baryon mass
$M_B(\mu)$ to be the location of the pole of $\langle\bar
B(x)B(y)\rangle$ in the broken phase gives
\begin{equation}
\label{bmass}
M_B(\mu) = \sinh^{-1}(m_B)-N\!\mu = 
\sinh^{-1}\biggl(4\bigl(2+\textstyle{3\over\gamma}\bigr)^{\textstyle{3\over
2}}\biggr) - 3\mu.
\end{equation}
In the symmetric phase the dynamical quark mass vanishes along with
the chiral order parameter.  The pole in $\langle\bar B(x)B(y)\rangle$
disappears, and so, as expected we no longer have stable nuclear
matter.  

\begin{figure}
\begin{center}
\leavevmode
\epsfxsize=4.3in
\epsfbox{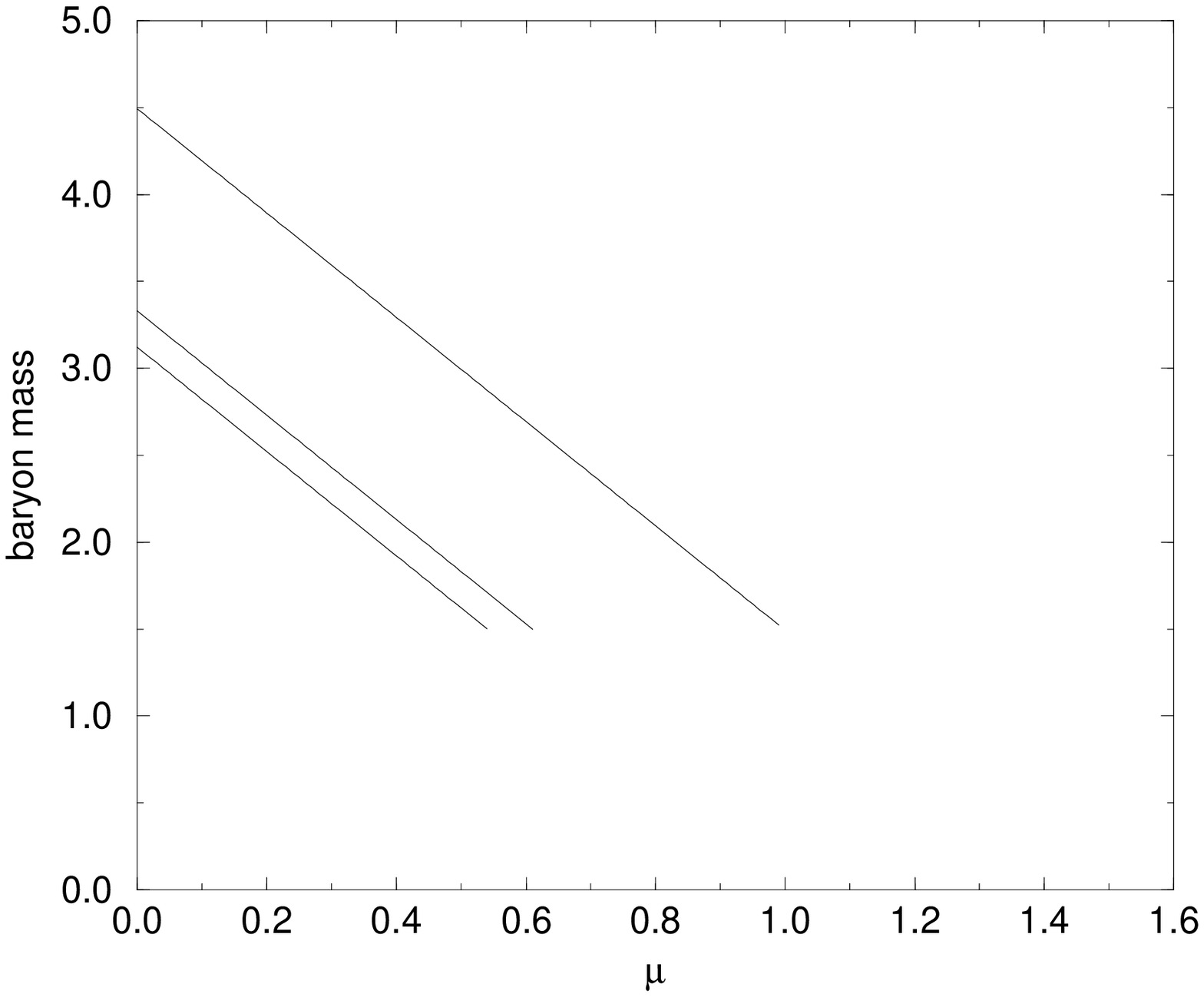}
\caption{Plots of the $\beta\!=\!0$ baryon mass as a
function of the chemical potential in the broken phase.  The three
lines are for $\gamma\!\to\!\infty$, $\gamma\!=\!10$, and
$\gamma\!=\!1$, from left to right.}
\label{fig:bmass}
\end{center}
\end{figure}

$M_B(\mu)$ is plotted in Fig.~\ref{fig:bmass} for various values of
$\gamma$.  Notice that the chiral symmetry restoration occurs
consistently at a value of $M_B(\mu_{\mbox{c}})\!=\!3/2$.  This value
can be traced back to the first two terms of $\bar{S}_{eff}$ in
(\ref{S_effk}), which represents the mean field energy density of the
chiral condensate.  At weaker gauge coupling, where the binding energy
is small, one would expect the phase transition to be closer to the
baryon mass threshold
$M_B(\mu_{\mbox{c}})\!=\!0$\cite{BarbourBDKMSW86}.  In ordinary QCD,
one finds that this energy difference does indeed decrease when
$1/{g^2}$ corrections are added to the calculations\cite{BilicDP92}.

The quark number density and quark energy density are given
by the following equations:
\begin{eqnarray}
\langle\psi^{\dagger}\!\psi\rangle 
&=& -\tr\biggl[{{\partial V_B\over\partial\mu}\over \Sigma^N 
+ V_B}\biggr] \nonumber \\
&=& 3\!\cdot\! 16\int_{-{\pi\over 2}}^{\pi\over 2}{\d^4 k \over (2\pi)^4}
{\textstyle{1\over
2}\bigl(\cos2k_0\sinh2N\mu + \i\sin2k_0\cosh2N\mu\bigr)
\over{\textstyle{1\over 2}\bigl(1-\cos2k_0\cosh2N\mu - \i
\sin2k_0\sinh2N\mu\bigr)+\displaystyle\sum_{\nu=1,2,3}\sin^2 k_{\nu} + m_B^2}},
\\
\langle\psi^{\dagger}\!\partial_0\psi\rangle
&=& \tr\biggl[{(V_B)^0\over \Sigma^N + V_B}\biggr] \nonumber \\
&=& 16\int_ {-{\pi\over 2}}^{\pi\over 2}{\d^4 k \over (2\pi)^4}
{\textstyle{1\over 2}\bigl(1-\cos2k_0\cosh2N\mu - \i
\sin2k_0\sinh2N\mu\bigr)\over
{\textstyle{1\over 2}\bigl(1-\cos2k_0\cosh2N\mu - \i
\sin2k_0\sinh2N\mu\bigr)+\displaystyle\sum_{\nu=1,2,3}\sin^2 k_{\nu} + m_B^2}}.
\end{eqnarray}
In the broken phase the baryon mass gap remains relatively large all
the way to the phase boundary.  Therefore, the baryon density is
strictly zero.  In the symmetric phase the binding energy of the quark
matter is very large at strong coupling.  Hence, the density jumps
directly to saturation as the chemical potential is increased through
the transition.  This is certainly an artifact of working at infinite
gauge coupling.  Similar behavior has been found in other infinite
coupling calculations\cite{BilicKR92}.

\section{Acknowledgments}
\begin{figure}
\begin{center}
\leavevmode
\epsfxsize=4.3in
\epsfbox{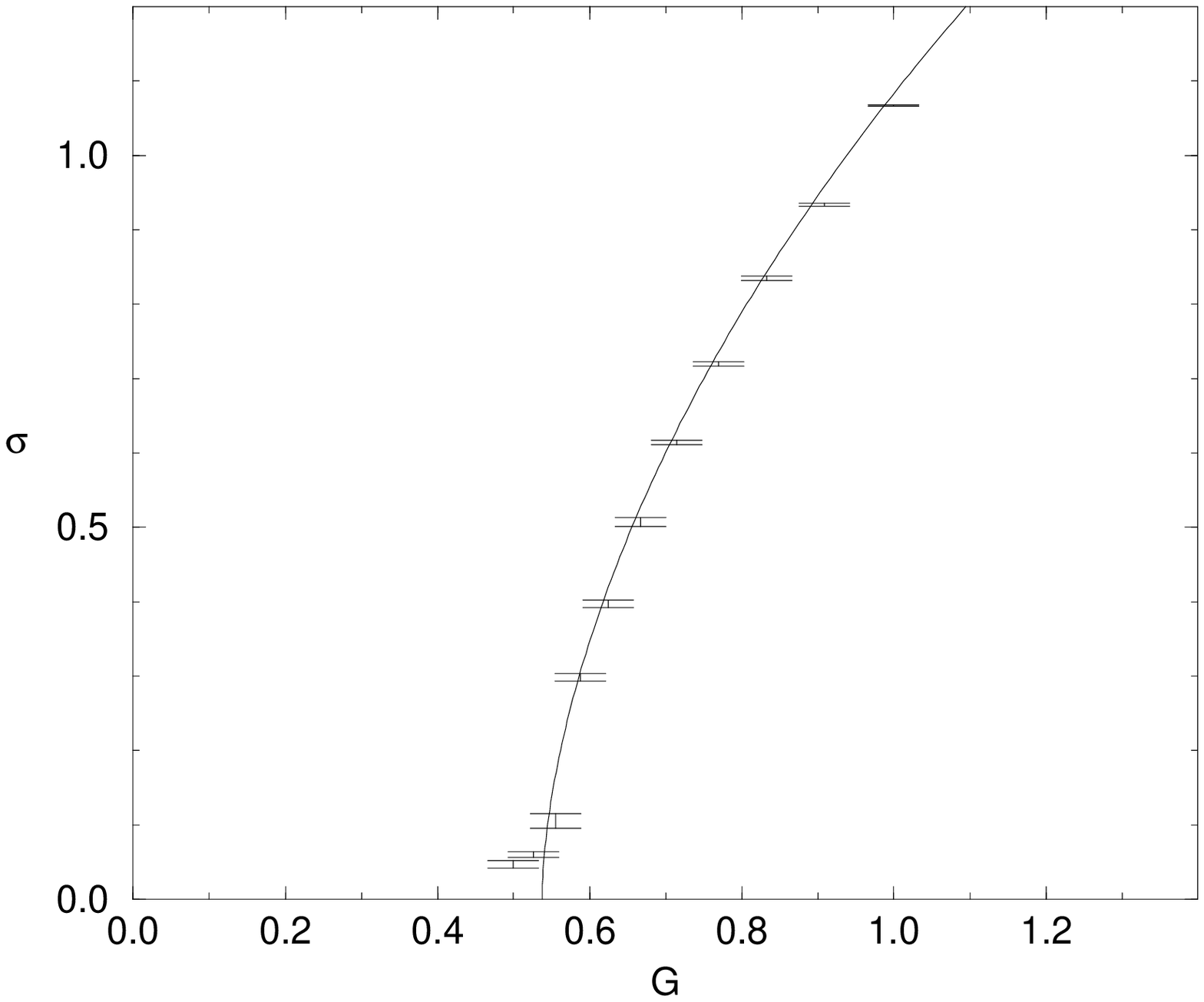}
\caption{The solid line shows $\bar\sigma$ vs. $G\!=\!1/\gamma$ as
given by the $\beta\!\to\!\infty$ gap equation at zero chemical
potential.  The data points are Monte Carlo measurements of
$\protect\sqrt{\sigma^2+\pi^2}$ at zero mass from
Ref.~\protect\cite{KogutLMB97}.}
\label{fig:NJLMCdata}
\end{center}
\end{figure}

\begin{figure}
\begin{center}
\leavevmode
\epsfxsize=4.3in
\epsfbox{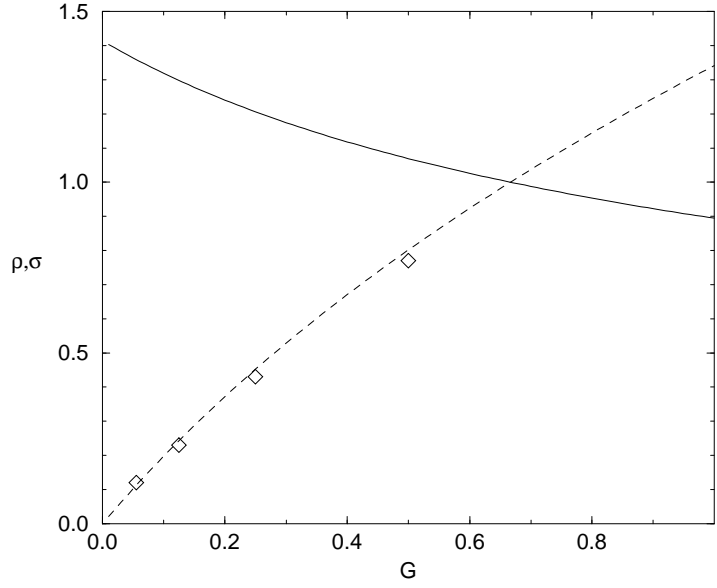}
\caption{Mean field values of $\bar\rho$ (solid line) and $\bar\sigma$
(dashed line) as a function of $G=1/\gamma$ in the broken phase, at
$\beta\!=\!0$. The diamonds are Monte Carlo values of $\sigma$ at
$\beta\!=\!0.5$ from Ref.~\protect\cite{KogutLMB97}.}
\label{fig:MCdata}
\end{center}
\end{figure}

The author would like to thank John B. Kogut for suggesting this
study and for providing the Monte Carlo data presented in
Figs.~\ref{fig:NJLMCdata} and \ref{fig:MCdata}.


\newpage

\begin{references}
\bibitem{Brower95}
R. C. Brower, Y. Shen and C.-I. Tan,
Int.~J. of Mod.~Phys.~{\bf C 6}, 743 (1995).
\bibitem{KogutS96}
 J.B. Kogut and D.K. Sinclair,
Nucl.~Phys.~{\bf B} (Proc.~Suppl.) 53, 272 (1997).
\bibitem{BarbourMK96}
 I.M. Barbour, S.E. Morrison, and J.B. Kogut,
 hep-lat/9612012.
\bibitem{WilsonK74}
 K.G. Wilson and J. Kogut, Phys.~Rep.~{\bf C 12}, 75 (1974).
\bibitem{DamgaardHK85}
 P.H. Damgaard, D. Hochberg, and N. Kawamoto,
Phys.~Lett.~{\bf 158B}, 239 (1985)
\bibitem{BarbourBDKMSW86}
 I. Barbour {\it et al.}, Nucl.~Phys.~{\bf
B275}, 296 (1986).
\bibitem{BilicKR92}
 N. Bili\'c, F. Karsch, and K. Redlich,
Phys.~Rev.~{\bf D 45}, 3228 (1992).
\bibitem{BilicDP92}
 N. Bili\'c, K. Demeterfi, and B. Petersson,
Nucl.~Phys.~{\bf B377}, 651 (1992).
\bibitem{BarbourKM96} I.M. Barbour, J.B. Kogut and S.E. Morrison,
Nucl.~Phys.~{\bf B} (Proc.~Suppl.) 53, 456 (1997).
\bibitem{quenched_QCD}
M.A. Stephanov, Phys.~Rev.~Lett. {\bf 76}, 4472 (1996).
\bibitem{KarschM89}
F. Karsch and K.-H. M\"utter,
Nucl.~Phys.~{\bf B313}, 541 (1989).
\bibitem{Gocksch88}
A. Gocksch,
Phys.~Rev.~Lett.~{\bf 61}, 2054 (1988).
\bibitem{Gibbs86}
P. Gibbs,
Phys.~Lett.~{\bf 172B}, 53 (1986);
Phys.~Lett.~{\bf 182B}, 369 (1986).
\bibitem{Barbour90}
I. Barbour, C. Davies, Z. Sabeur, Phys.~Lett.~{\bf B 215}, 567 (1988); 
I. Barbour and Z. Sabeur, Nucl.~Phys.~{\bf B342}, 269 (1990); 
I. M. Barbour and A.J. Bell, Nucl.~Phys.~{\bf B372}, 385 (1992).
\bibitem{staggered}
J. Kogut and L. Susskind,
Phys.~Rev.~{\bf D 16}, 395 (1975);
T. Banks, J. Kogut, and L. Susskind, Phys.~Rev.~{\bf D 13}, 1043 (1976);
L. Susskind,
Phys.~Rev.~{\bf D 16}, 3031 (1977); 
See I. Montvay and G. M\"unster,
{\it Quantum Fields on a Lattice} (Cambridge University Press, 1994),
chapter 4, for an introduction.
\bibitem{HasenfratzK83}
 P. Hasenfratz and F. Karsch, Phys.~Lett.~{\bf 125B}, 308 (1983).
\bibitem{KawamotoS81}
N. Kawamoto and J. Smit,
Nucl.~Phys.~{\bf B192}, 100 (1981).
\bibitem{JolicoeurMP86}
 T. Jolicoeur, A. Morel and B. Petersson,
Nucl.~Phys.~{\bf B274}, 225 (1986).
\bibitem{HandsKK93}
 S. Hands, A. Koci\'c and J.B. Kogut, Ann.~Phys.
{\bf224}, 29 (1993)
\bibitem{NJL}
 See S.P. Klevansky, Rev. Mod. Phys.~{\bf 64}, 649 (1992)
for a review.
\bibitem{HandsKK95}
 S. Hands, S. Kim and J.B. Kogut,
Nucl.~Phys.~{\bf B442}, 364 (1995).
\bibitem{KogutLMB97}
J. Kogut, M.-P. Lombardo, I. Barbour, and S. Morrison (unpublished data). 
\bibitem{KimKK94}
S. Kim, A. Koci\'c and J. Kogut,
Nucl.~Phys.~{\bf B429}, 407 (1994).
\bibitem{Kluberg-SternMP83}
 H. Kluberg-Stern, A. Morel and B.
Petersson, Nucl.~Phys~{\bf B215}, 527 (1983); {\bf
B220}, 447 (1983).
\end{references}
\end{document}